%% file: Template.tex
\documentclass{article}
\usepackage{spconf,amsmath,graphicx,amsfonts}
\usepackage{stfloats}

\title{Deep3DSketch: 3D modeling from Free-hand Sketches with View- and Structural-Aware Adversarial Training}
\name{Tianrun Chen$^{1}$, Chenglong Fu$^{2}$, Lanyun Zhu$^{3}$, Papa Mao$^{4}$, Jia Zhang$^{4}$, Ying Zang$^{2*}$, Lingyun Sun$^{1}$}

\address{\small $^{1}$College of Computer Science and Technology, Zhejiang University, Hangzhou, Zhejiang, P.R. China.\\ \small $^{2}$School of Information Engineering, Huzhou University, Huzhou, Zhejiang, P.R. China.\\ \small $^{3}$Information Systems Technology
and Design Pillar, Singapore University of Technology and Design, Singapore.\\ \small $^{4}$Mafu Laboratory, Moxin (Huzhou) Tech. Co., LTD, Huzhou, Zhejiang, P.R. China.}

\begin{document}
%
\maketitle
\begin{abstract}
This work aims to investigate the problem of 3D modeling using single free-hand sketches, which is one of the most natural ways we humans express ideas. Although sketch-based 3D modeling can drastically make the 3D modeling process more accessible, the sparsity and ambiguity of sketches bring significant challenges for creating high-fidelity 3D models that reflect the creators' ideas. In this work, we propose a view- and structural-aware deep learning approach, \textit{Deep3DSketch}, which tackles the ambiguity and fully uses sparse information of sketches, emphasizing the structural information. Specifically, we introduced random pose sampling on both 3D shapes and 2D silhouettes, and an adversarial training scheme with an effective progressive discriminator to facilitate learning of the shape structures. Extensive experiments demonstrated the effectiveness of our approach, which outperforms existing methods -- with state-of-the-art (SOTA) performance on both synthetic and real datasets. 
\end{abstract}
\renewcommand{\thefootnote}{}
\footnotetext{* 02750@zjhu.edu.cn}
\section{Introduction}
\label{sec:intro}

The rapid development of portable displays and AR/VR brings tremendous demands for 3D content \cite{wang2020vr}. Computer-Aided Design (CAD) methods require creators to master sophisticated CAD software commands (\textit{commands knowledge}) and to be able to parse a shape into sequential commands (\textit{strategic knowledge}), which restricts its application in expert users \cite{bhavnani1999strategic, chester2007teaching}. The restrictions call for the need for alternative methods to open the door to 3D modeling for the masses. In recent years, sketch-based 3D modeling has been recognized as a potential solution, as sketches are one of the most natural ways we humans express ideas. While many works have proposed to perform 3D modeling using sketches, Most existing works either require precise line drawings from multiple views or apply step-by-step workflow with \textit{strategic knowledge} required \cite{ cohen1999interface, deng2020interactive}, which is not friendly for novice users. Other work use template primitives or retrieval-based approaches \cite{chen2003visual, wang2015sketch}, but lack the customizability.

To mitigate the research gap, we aim to use only one single sketch as the input to generate a complete and high-fidelity 3D model. The approach is designed to fully exploit the human sketches to develop an intuitive and fast 3D modeling approach -- generating a high-fidelity 3D model that represents the creators' intention. 

However, generating a 3D model from a single sketch is non-trivial. The sparsity and ambiguity of sketches bring significant challenges. Specifically, sketches are sparse because they have only a single view, are mostly abstract, lack fine boundary information when drawing by humans, and, more critically, lack the texture information for depth estimation. This brings large uncertainty when learning 3D shapes. The abstract boundary also makes it hard to interpret, as the same set of strokes may lead to different interpretations in the 3D world, which leads to ambiguity. Existing works \cite{guillard2021sketch2mesh,zhang2021sketch2model} have demonstrated that deploying a widely-used auto-encoder as the backbone of the network can only obtain coarse prediction, but is unable to obtain the fine-grained 3D structures. 

Facing the challenges, we present our \textbf{Deep3DSketch}, a novel and more effective sketch-based modeling approach, which can obtain 3D shapes with fine-grained and reasonable 3D structures. Specifically, we first explicitly learn the view information and use it to condition the generation process to resolve the ambiguity. We then perform random pose sampling to force learning of realistic and high-fidelity 3D shapes independent from the viewpoint. The disentanglement is similar to disentangling "where" and "what" \cite{zhu2021and}. We also introduce an adversarial training scheme with an effective progressive discriminator that is aware of the geometric structure of the objects via cross-view silhouettes of the 3D model. The discriminator alleviates the uncertainty from the sparsity through more visual clues from different viewpoints, leading to better optimization results. Extensive experiments demonstrated the effectiveness of our approach for generating 3D models with higher fidelity, achieving state-of-the-art (SOTA) performance on both synthetic and real datasets. 

\section{Method}

\subsection{Preliminary}

Given the input binary sketch $I  \in \left \{ 0,1 \right \}^{W\times H}$ , the goal of the network $G$ is to obtain a mesh $M_\Theta =(V_\Theta,  F_\Theta)$, in which $V_\Theta$ and $F_\Theta$ represents the mesh vertices and facets, and the rendered silhouette $S_\Theta:$
$\mathbb{R}$$^3$ 
$\rightarrow$ 
$\mathbb{R}$$^2$ 
of $M_\Theta $ matches with the information from the input sketch $I$. We use commonly-used encoder-decoder structure as the backbone, an Encoder $E$ is used to obtain a compressed shape code $z_s$ and a Decoder $D$ manipulate $z_s$ to calculate the vertex offsets of a template mesh and deforms it to get the output mesh $M_\Theta = D(z_s)$.
\begin{figure*}[t]
\centering
\includegraphics[width=0.9\linewidth]{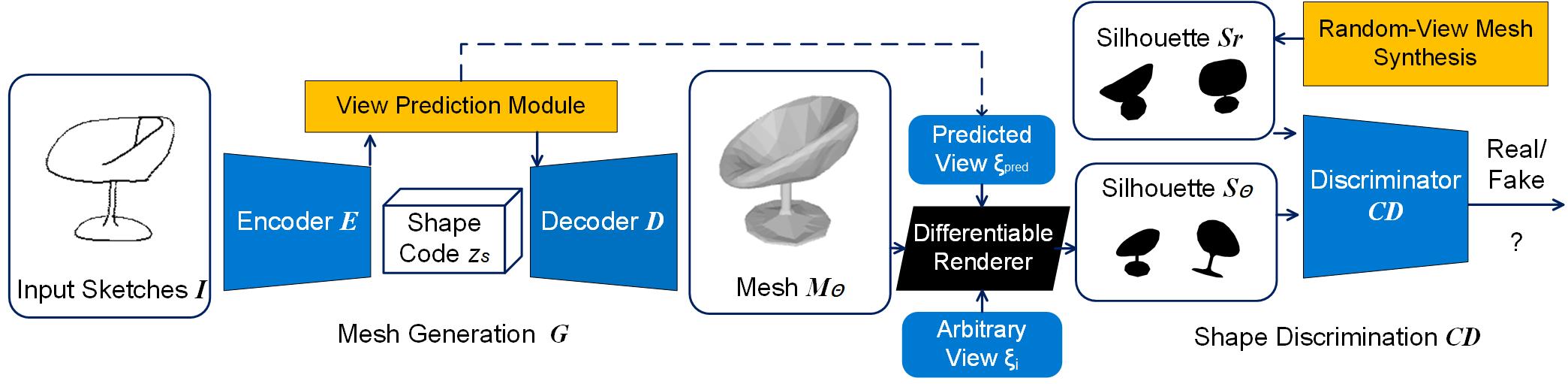}
\caption{{\textbf{The structure of Deep3DSketch.} View- and structural- aware sketch-based 3D modeling with adversarial training.}}
\label{fig:0}
\end{figure*}

\subsection{View-Aware 3D Model Generation}

We first introduce extra clues -- view information, which can tackle the challenge of ambiguity \cite{zhang2021sketch2model}. As we humans use viewpoint clues to recognize and interpret 3D objects, viewpoint clues are important in single sketch-based 3D modeling, especially in resolving ambiguity. Therefore, we explicitly learns the viewpoint and use the viewpoint information to condition the generation process. We first let the encoder $E$ produce another latent code $z_l$ and input it to the viewpoint prediction module. We implement two fully-connected layers to produce the viewpoint estimation $\xi_{pred}$, represented by an Euler angle. The viewpoint prediction module is optimized in a fully-supervised manner, with the input of the ground truth viewpoint and supervised by a viewpoint prediction loss $\mathcal{L}_{v}$, which adopted MSE loss for predicted and ground truth camera pose, defined as:
\begin{align}
\mathcal{L}_{v}=\|V-\hat{V}\|_{2}=\left\|V-D_{v}\left(z_{v}\right)\right\|_{2}
\end{align}
The output viewpoint prediction $\xi_{pred}$ is fed into a differentiable renderer to render silhouette at the given viewpoint for supervision. Specially, we use the mIoU Loss $\mathcal{L}_{iou}$ to measure the similarity between the rendered silhouette $S_1$ 
 and the silhouette of the input sketch $S_2$:
\begin{align}
\mathcal{L}_{i o u}\left(S_{1}, S_{2}\right)=1-\frac{\left\|S_{1} \otimes S_{2}\right\|_{1}}{\left\|S_{1} \oplus S_{2}-S_{1} \otimes S_{2}\right\|_{1}}
\end{align}
For computational efficiency, we progressively increase the resolutions of silhouettes, forming the multi-scale mIoU loss $\mathcal{L}_{sp}$, which is represented as 
\begin{align}
\mathcal{L}_{s p}=\sum_{i=1}^{N} \lambda_{s_{i}} \mathcal{L}_{iou}^{i}
\label{lsp}
\end{align}

The predicted viewpoint is also used to guide the generation process. We fed the viewpoint into another two fully-connected layers to produce a view-aware vector representation $z_v$, and input both $z_v$ and  $z_s$ to the Decoder $D$ to produce the $M_\Theta$. 

To further condition the generation process with viewpoint constraints, we add a Random-View Mesh Synthesis branch, in which a random viewpoint $\xi_{random}$ is obtained and a mesh $M_\Theta r$ is generated following the same manner as mesh generation with $\xi_{pred}$. The generated $M_\Theta r$ with random (fake) viewpoint constraint is regarded as the fake sample, while the generated mesh $M_\Theta$ is regarded as the real sample. They together feed into a Shape Discriminator $CD$, to force the neural network generate meshes under the view-constraint. 
\subsection{Structural-Aware 3D Model Generation}
So far, the supervision of the mesh generation fidelity is from a single rendered silhouette of generated mesh with a given viewpoint as the input. With only 2D input as the supervision, our goal is, however, to obtain complete 3D shapes with fine-grained structural information. A single sketch and the corresponding silhouette can only represent the information at that given viewpoint, but lacks the information from other viewpoints, thus making it hard to obtain detailed structural information. The sparsity of the sketch contributed to the difficulty of obtaining fine-grained structures. Therefore, we propose to have multiple random-view silhouettes. The random pose sampling aims to force the network learns to generate reasonable 3D fine-structured shapes independent from the viewpoints. In addition, as many previous works investigated in the realm of shape-from-silhouette, the proposed multi-view silhouettes contain valuable geometric information about the 3D object~\cite{gadelha2019shape,hu2018structure,zheng2009robust}. In practice, we randomly sample $N$ camera poses $\xi_{1...N}$ from camera pose distribution $ p_{\xi} $. We use a differentiable renderer to render the silhouettes $S_{1...N}$ from the mesh $M$ and render the silhouettes $S_r\left \{1...N \right \} $ from the mesh $M_r$. The differentiable renderer $R$ is shown in \cite{liu2019soft}. By introducing the $S_r\left \{1...N \right \} $ , the network is aware of the geometric structure of the objects in cross-view silhouettes when generating the 3D objects, and the discriminator helps to resolve the challenge from the sparsity of sketches by offering more visual clues. The disentanglement is very similar to disentangling "where" and "what" principles in generative models \cite{zhu2021and}, which is proven to be effective in our tasks. 


In addition, to fully capture the structural information of the rendered silhouettes, we apply a convolutional progressive growing discriminator $CD$. Following \cite{karras2017progressive}, our discriminator is trained with increasing image resolution and incrementally added new layers to handle the higher resolutions and discriminate fine details. We discovered that such convolutional discriminator design is more effective in capturing local and global structural information to facilitate the generation of high-fidelity 3D shapes, compared to MLP-enabled discriminator for 3D objects. In training, non-saturating GAN loss with R1 regularization is used \cite{mescheder2018training} for better convergence:

\begin{align}
\begin{split}
\mathcal{L}_{sd} &=\mathbf{E}_{\mathbf{z_v} \sim p_{z_v}, \xi \sim p_{\xi}}\left[f\left(CD_{\theta_{D}}\left(R(M, \xi)\right)\right)\right] \\
&+\mathbf{E}_{\mathbf{z_{vr}} \sim p_{z_{vr}}, \xi \sim p_{\xi}}\left[f\left(-CD_{\theta_{D}}(R(M_r, \xi))\right)\right] \label{gan}
\end{split}\\ 
&\textit { where } f(u)=-\log (1+\exp (-u))
\end{align}

\subsection{Domain Adaptation}
Due to the lack of large amount of ground truth 3D models and the corresponding 2D sketches, we use synthetic data for training and testing at real-world data, in which the domain gap exists. To make our network generalizable to real hand-draw datasets, we applied domain adaptation (DA) technique and introduce the DA loss $\mathcal{L}_{dd} $, as the same in~\cite{zhang2021sketch2model}.
\begin{figure*}[b]
\centering
\includegraphics[width=\textwidth]{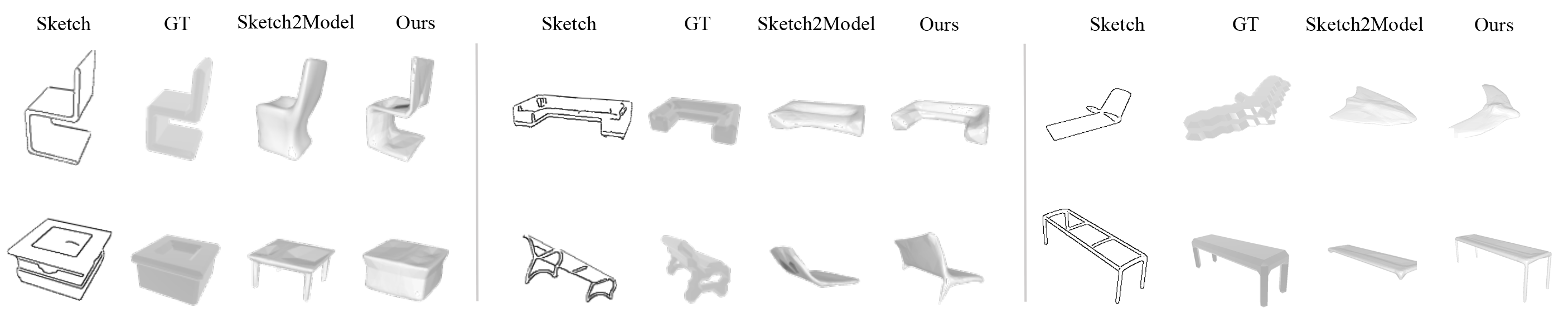}
\caption{{\textbf{Qualitative evaluation with existing state-of-the-art.} The visualization demonstrated our method's capability of synthesizing higher fidelity 3D structures.}}
\label{fig:2}
\end{figure*}
\subsection{Training Details}
\textbf{Loss Function. }
To make meshes more realistic with higher visual quality, we also use flatten loss and Laplacian smooth loss in~\cite{zhang2021sketch2model, kato2018neural, liu2019soft} 
, represented by $\mathcal{L}_{r}$. 
The overall loss function $\mathcal{L}$ is calculated as the weighted sum of five components:
\begin{align}
\mathcal{L} =  \mathcal{L}_{sp} + \mathcal{L}_{r}+ \lambda_v \mathcal{L}_v  + \lambda_{sd} \mathcal{L}_{sd} + \lambda_{dd} \mathcal{L}_{dd} 
\label{loss}
\end{align}
\textbf{Implementation Details.} 
We use ResNet-18 \cite{he2016deep} as the encoder for image feature extraction. The extracted 512-dim feature goes through 2 linear layers with L2-normalization and generates a 512-dim shape code $z_s$ and a 512-dim view code $z_v$. The rendering module is SoftRas \cite{liu2019soft}, the number of views $N=3$. Each 3D object is placed with 0 in evaluation and 0 in azimuth angle in the canonical view, with a fixed distance to the camera. We use Adam optimizer with an initial learning rate of 1e-4, and multiply by 0.3 for every 800 epochs. Betas equal 0.9 and 0.999. The total training epochs equal 2000. The model is trained individually with each class. $\lambda_{r}$, $\lambda_{sd}$, and $\lambda_{dd}$ in Equation. \ref{loss} equal to 0.1, $\lambda_{v}$ and $\lambda_{vr}$, equal to 10. 
When evaluating with the ShapeNet-Sketch dataset, we use domain adaptation on 7 of the classes, which have sufficient amount of sketches in the Sketchy dataset \cite{sangkloy2016sketchy} and Tu-Berlin dataset \cite{eitz2012humans}. The domain adaptation is performed by concatenating the average pooling and max pooling results of the image feature map as input, as in \cite{woo2018cbam}. 

\section{Experiments}

\subsection{Datasets}

Training the model requires large-scale sketch data with the corresponding 3D models, which is rare in the public domain. Following \cite{zhang2021sketch2model}, we use the synthetic data \textit{\textbf{ShapeNet-synthetic }}for training and testing, and the real-world data \textit{\textbf{ShapeNet-Sketch}} to evaluate the method in the wild. ShapeNet-synthetic is the edge map extracted by a canny edge detector provided by Kar et al. \cite{kar2017learning}. It contains 13 categories of 3D objects from ShapeNet. The ShapeNet-Sketch is a dataset collected from real-human drawings. Volunteers with varied drawing skills are asked to draw objects based on the rendered images of 3D objects, with a total number of 1300 sketches and their corresponding 3D shapes.

\subsection{Results}
\noindent \textbf{The ShapeNet-Synthetic Dataset.  }
We first evaluate the performance on the dataset with the ground truth 3D model. Meshes with the predicted viewpoint (Pred Pos) and the ground truth viewpoint (GT Pos) are trained and evaluated, respectively. We apply common-used 3D reconstruction metrics -- voxel IoU to measure the fidelity. The result is shown in Table \ref{table:table1}. Our method achieves state-of-the-art (SOTA) performance in every category. The quantitative evaluation of our method compared with existing state-of-the-art in Figure  \ref{fig:2} further demonstrated the effectiveness of our approach to reconstructing models with higher fidelity in structure.
\input{table/table1}

\input{table/table2}

\noindent \textbf{The ShapeNet-Sketch Dataset.  }
We further evaluate the performance of real-world human drawings. We train the model on ShapeNet-Synthetic dataset and use ShapeNet-Sketch dataset for evaluation. As shown in Table \ref{table:table2}, our model outperforms the existing state-of-the-art methods in most categories. Our method outperforms the existing method in some categories even without Domain Adaptation (DA). 

\subsection{Ablation Study}

\begin{figure}[h]
\includegraphics[width=0.5\textwidth]{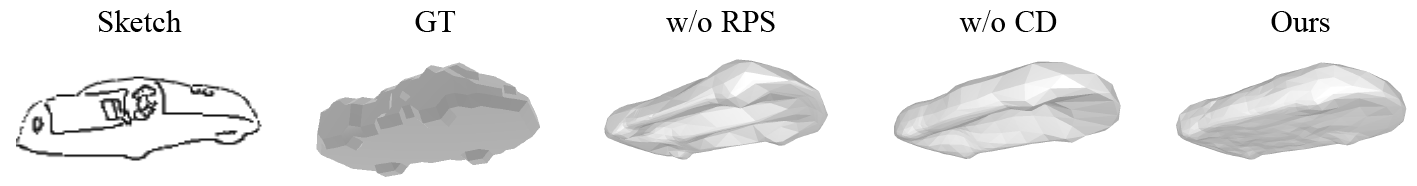}
\caption{{\textbf{Visualization of Ablation.} The network generates unwanted structures w/o RPS and unrealistic structure w/o CD. }}
\label{fig:3}
\end{figure}

To show the effectiveness of our proposed method, we conducted the ablation study that removes Random Pose Sampling (RPS) for view-awareness. We also remove the progressive Convolutional Discriminator (CD) and use an MLP-based discriminator as in \cite{zhang2021sketch2model}. Our quantitative result (Table \ref{table:table3}) and qualitative example (Figure \ref{fig:3}) shows removing the RPS and CD will be detrimental to the performance.

\input{table/table3}

\section{Conclusion}

We propose \textit{Deep3DSketch}, a novel 3D modeling approach that generates 3D models with only a single sketch. For high-fidelity 3D modeling, we disentangle the learning of view and structural learning. We first condition the generation on an explicitly learned viewpoint, then use random pose sampling for viewpoint-independent learning of shapes and fully exploit the geometric information in cross-view silhouettes. We also introduce a progressive convolutional discriminator to better capture the structural information of a 3D mesh at local and global levels. With the alleviated ambiguity and sparsity, we have shown state-of-the-art (SOTA) performance on both real and synthetic data. We believe our method has great potential to revolutionize future 3D modeling pipelines.  

\section{ACKNOWLEDGEMENT}
This work is supported by National Key R$\&$D Program of China (2018AAA0100703). We thank Zejian Li for advice.

\bibliographystyle{IEEEbib}
\bibliography{strings,refs}

\end{document}

%% file: table/table1.tex
\begin{table*}[ht]
\begin{center}
\caption{\textbf{The quantitative evaluation of ShapeNet-Synthetic dataset.}}
\scalebox{0.71}{

\begin{tabular}{|c|c|c|c|c|c|c|c|c|c|c|c|c|c|c|c|}

\hline
\multicolumn{15}{|c|}{Shapenet-synthetic (Voxel Iou $\uparrow)$} \\
\hline
 & car & sofa & airplane & bench & display & chair & table  & telephone & cabinet & loudspeaker & watercraft & lamp & rifile & mean\\
\hline
Retrieval & 0.667 & 0.483 & 0.513 & 0.38  & 0.385 & 0.346 & 0.311 & 0.622 & 0.518 & 0.468 & 0.422 & 0.325 & 0.475 & 0.455\\
\hline
Auto-encoder & 0.769 & 0.613 & 0.576 & 0.467 & 0.541 & 0.496 & 0.512  & 0.706 & 0.663 & 0.629 & 0.556 & 0.431 & 0.605 & 0.582\\
\hline
Sketch2Model (GT Pos)  & 0.751 & 0.622 & 0.624 & 0.481 & 0.604 & 0.522 & 0.478 & 0.719 & 0.701 & 0.641 & 0.586 & 0.472 & 0.612 & 0.601 \\
Sketch2Model (Pred Pos) & 0.746 & 0.620 & 0.618 & 0.477 & 0.550 & 0.515 & 0.470 & 0.673 & 0.667 & 0.624 & 0.569 & 0.463 & 0.606 & 0.584 \\
\hline
\textbf{Ours (GT Pos)} & \textbf{0.796} & \textbf{0.651} & \textbf{0.644} & \textbf{0.500} & \textbf{0.612}	& \textbf{0.544} & \textbf{0.518} & \textbf{0.738} & \textbf{0.705} & \textbf{0.651} & \textbf{0.595} & \textbf{0.469} & \textbf{0.619} & \textbf{0.618} \\
\textbf{Ours (Pred Pos)} & 0.793 & 0.649 & 0.641 & \textbf{0.500} & 0.583 & 0.541 & 0.504 & 0.680 & 0.683 & 0.623 & 0.580 & 0.465 & \textbf{0.619} & 0.604\\
\hline
\end{tabular}}
\end{center}
\vspace{-0.65cm}
\label{table:table1}

\end{table*}

%% file: table/table2.tex

\begin{table*}[ht]
\begin{center}
\caption{\textbf{The quantitative evaluation of ShapeNet-Sketch dataset.}}
\scalebox{0.66}{

\begin{tabular}{|c|c|c|c|c|c|c|c|c|c|c|c|c|c|c|c|}

\hline
\multicolumn{15}{|c|}{Shapenet-sketch (Voxel Iou $\uparrow$)} \\
\hline
 & car & sofa & airplane & bench & display & chair & table & telephone & cabinet & loudspeaker & watercraft & lamp & rifile & mean\\
\hline
Retrieval & 0.626 & 0.431 & 0.411 & 0.219 & 0.338 & 0.238 & 0.232 & 0.536 & 0.431 & 0.365 & 0.369 & 0.223 & 0.413 & 0.370\\
\hline
Auto-encoder & 0.648 & 0.534 & 0.469 & 0.347 & 0.472 & 0.361 & 0.359 & 0.537 & 0.534 & 0.533 & 0.456 & 0.328 & 0.541 & 0.372 \\
\hline
Sketch2Model (GT Pos)  & 0.659 & 0.534 & 0.487 & 0.366 & 0.479 & 0.393 & 0.357 & 0.554 & \textbf{0.568} & 0.526 & 0.450 & \textbf{0.338} & 0.534 & 0.483 \\
Sketch2Model (Pred Pos) & 0.649 & 0.528 & 0.479 & 0.357 & 0.435 & 0.383 & 0.361 & 0.551 & 0.547 & 0.544 & 0.466 & 0.336 & 0.510 & 0.470\\
\hline
Sketch2Model + DA (GT Pos) & 0.679  & \textbf{0.548} & \textbf{0.526} & \textbf{0.367} & - & \textbf{0.398} & 0.357 & - & - & - & - & - & 0.535 & 0.489\\
Sketch2Model + DA (Pred Pos) & 0.659 & 0.533 & 0.515 & 0.362 & & 0.385 & 0.360  &  &  &  &  &  & 0.511 & 0.475 \\
\hline
\textbf{Ours (GT Pos)} & 0.695 & 0.528 & 0.502 & 0.364 & \textbf{0.493} & 0.389 & 0.370 & \textbf{0.574} & 0.563 & \textbf{0.538} & \textbf{0.477} & 0.334 & 0.535 & 0.489 \\
\textbf{Ours (Pred Pos)} & 0.683 & 0.523 & 0.502 & 0.364 & 0.493 & 0.389 & \textbf{0.370} & 0.527 & 0.549 & 0.509 & 0.468 & 0.331 & 0.535 & 0.476 \\
\hline
\textbf{Ours + DA (GT Pos)}   & \textbf{0.699} & 0.538 & 0.517 & 0.362 & - & 0.390 & 0.360 & - & - & - & - & - & \textbf{0.545} & \textbf{0.491} \\
\textbf{Ours + DA (Pred Pos)} & 0.692 & 0.532 & 0.515 & 0.360 & - & 0.382 & 0.346  &  &  &  &  &  & \textbf{0.545} & 0.477\\
\hline

\end{tabular}}
\end{center}

\label{table:table2}
\vspace{-0.65cm}
\end{table*}

%% file: table/table3.tex
\begin{table}[h]
\caption{\textbf{Quantitative evaluation of ablation study.} }
\begin{center}
\scalebox{0.49}{
\begin{tabular}{c c || c c c c c c c }
\hline
\multicolumn{9}{c}{Ablation Study. (Numbers inside and outside the parenthesis are IoU on Pred View and GT View, respectively)} \\
\hline
\hline
RPS & CD & car & sofa & airplane & bench & display & chair & table \\
\hline
\hline
 & & 0.747 (0.753) & 0.624 (0.643) & 0.557 (0.565) & 0.345 (0.460) & 0.457 (0.577) & 0.499 (0.508) & 0.406 (0.427) \\
$\surd$ & & 0.782 (0.773) & 0.641 (0.639) & \textbf{0.644} (0.639) & 0.461 (0.485) & 0.597 (0.540) & 0.543 (0.538) & 0.512 (0.477) \\
$\surd$ & $\surd$ & \textbf{0.796} (\textbf{0.793}) & \textbf{0.651} (\textbf{0.649}) & \textbf{0.644} (\textbf{0.641}) & \textbf{0.500} (\textbf{0.500}) & \textbf{0.612} (\textbf{0.583}) & \textbf{0.544} (\textbf{0.541}) & \textbf{0.518} (\textbf{0.504}) \\
\hline
RPS & CD & telephone & cabinet & loudspeaker & watercraft & lamp & rifile & mean \\
\hline
\hline
 & & 0.522 (\textbf{0.705}) & 0.597 (0.579) & 0.584 (0.614) & 0.574 (0.575) & 0.290 (0.421) & 0.500 (0.576) & 0.516 (0.569) \\
$\surd$ & & 0.734 (0.673) & 0.696 (0.645) & 0.636 (0.599) & 0.585 (0.553) & \textbf{0.478} (\textbf{0.471}) & \textbf{0.619} (\textbf{0.627}) & 0.608 (0.588) \\
$\surd$ & $\surd$ & \textbf{0.738} (0.680) & \textbf{0.705} (\textbf{0.683}) & \textbf{0.651} (\textbf{0.623}) & \textbf{0.595} (\textbf{0.580}) & 0.469 (0.465) & \textbf{0.619} (0.619) & \textbf{0.618} (\textbf{0.604}) \\
\hline
\end{tabular}}
\end{center}
\vspace{-0.3cm}
\label{table:table3}

\end{table}